\documentclass[journal]{IEEEtran}
\usepackage{cite}
\usepackage{rotating} 
\usepackage{amsmath}
\usepackage{url}
\usepackage{tikz}
\usepackage{mathtools}

\newcommand\submittedtext{%
  \footnotesize This work has been submitted to the IEEE for possible publication. Copyright may be transferred without notice, after which this version may no longer be accessible.}

\newcommand\submittednotice{%
\begin{tikzpicture}[remember picture,overlay]
\node[anchor=south,yshift=10pt] at (current page.south) {\fbox{\parbox{\dimexpr0.65\textwidth-\fboxsep-\fboxrule\relax}{\submittedtext}}};
\end{tikzpicture}%
}

\ifCLASSINFOpdf
\else
\fi
\begin{document}
%
\title{Spatial-to-Spectral Harmonic-Modulated Arrays for 6G Multi-Beam MIMO}

\author{Jose~Guajardo, Ali~Niknejad}

\maketitle
\submittednotice

\begin{abstract}
This article presents an overview and analysis of spatial-to-spectral harmonic-modulated arrays (SHAs). Compared to traditional analog or digital beamforming arrays, SHAs enable concurrent multi-beamforming without requiring substantial  hardware replication. SHAs replace the need for hardware replication with frequency-domain multiplexing. Furthermore, SHAs have the potential to become key contributors to future 6G networks by enabling scalable multi-user communications, joint communication and sensing, and spatial interference mitigation. In addition, an analysis of the SHA's harmonic-modulation waveform and its effects on gain, noise and bandwidth is presented. A comb-like modulation waveform for SHAs that minimizes spectral inefficiency is proposed. Further, an analysis of the SHA's capability to independently steer multiple beams is presented. This capability is quantified in terms of the SHA's spatial-to-spectral degrees of freedom. Lastly, this work introduces a novel SHA architecture that provides three spatial-to-spectral degrees of freedom with minimal hardware replication.
\end{abstract}


\section{Motivation}
\label{section:motivation}
Multi-input multi-output (MIMO) systems have become a key technology for emerging wireless communication networks, including 6G. Multi-beam MIMO (MB-MIMO) systems significantly improve wireless channel capacity and enable key functionalities like rapid angle estimation or localization for communication systems \cite{garcia_direct_2017}. Moreover, MB-MIMO is also particularly beneficial for radar systems where it enables enhanced multi-target detection and tracking capabilities \cite{feger_77-ghz_2009,pfeffer_fmcw_2013}.

MB-MIMO systems are often implemented using analog beamforming arrays, where element-level phase shifters and subsequent power-combining networks synthesize a beam toward a desired direction \cite{park_41_2020,pellerano_97_2019,liang_2525-ghz_2025,sadhu_28-ghz_2017,shahramian_fully_2019,li_eight-element_2022,elkhouly_fully_2022}. Digital beamforming arrays are an alternative  where beam synthesis instead occurs in the digital backend \cite{lu_16-element_2021,farid_fully_2022,dosluoglu_reconfigurable_2022,jang_true_2018}. Both analog and digital architectures encounter significant challenges when scaling array size and number of beams, making MB-MIMO systems difficult to realize in practice. On one hand, a $K$-beam analog beamforming array requires $K$ independent beamformers leading to a significant power, area and cost overhead due to hardware replication. On the other hand, an $N$-channel digital beamforming array requires a complete RF-to-digital signal chain for each of the $N$ channels and, thus, also incurs significant overhead as the array size grows. Hybrid beamforming aims to strike a balance between analog and digital beamforming \cite{naviasky_71--86-ghz_2021,mondal_2530_2018,mondal_dual-band_2022}. However, it still inherits some of the the power, area and cost penalties due to the inherent hardware replication in these architectures.

\begin{figure}
    \centering
    \includegraphics[width=1\linewidth]{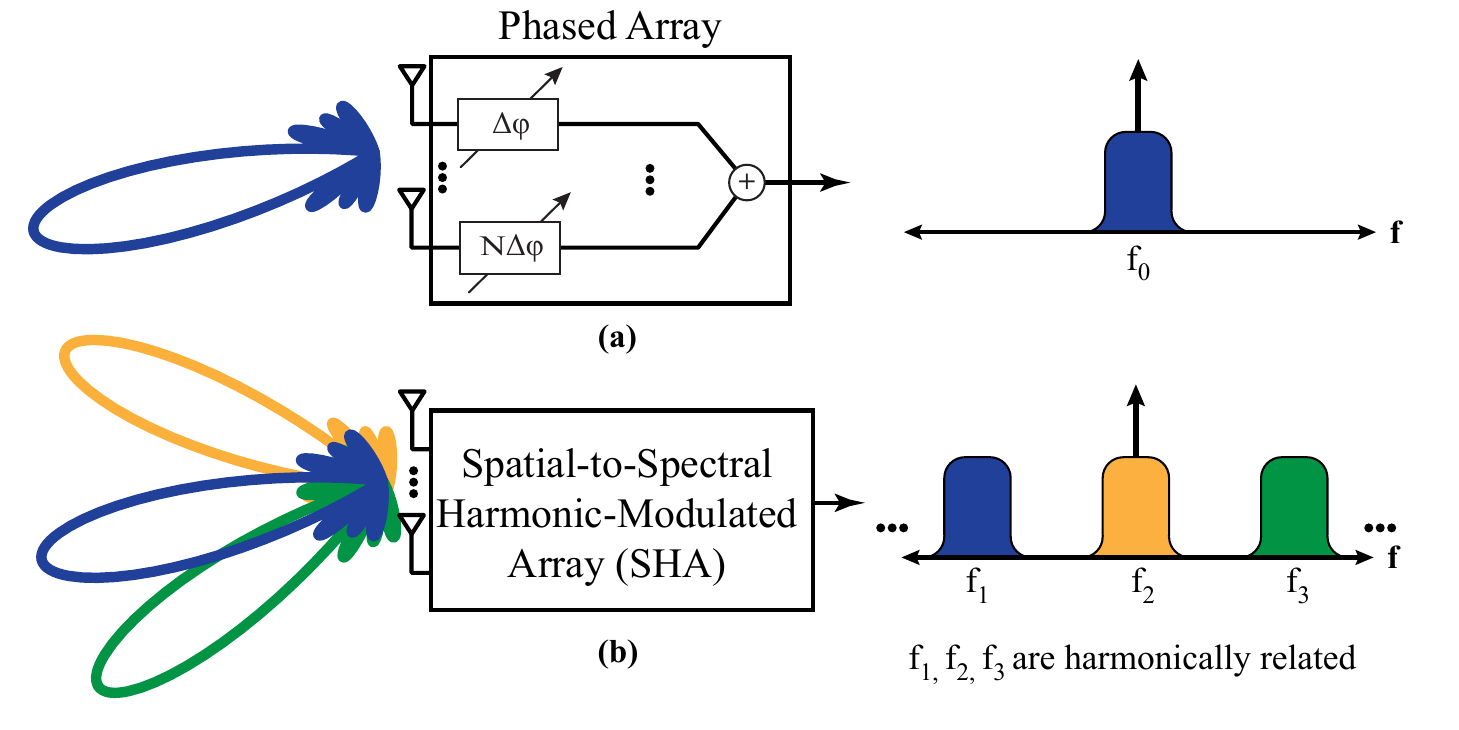}
    \caption{(a) A phased array typically forms a single beam and transmits/receives a signal centered at a single frequency $f_0$. (b) An SHA forms multiple beams at several frequencies $f_1, f_2, f_3 ...$ for transmit or receive operation.}
    \label{fig:spatialspectralintro}
\end{figure}

As illustrated in Fig.~\ref{fig:spatialspectralintro}, a spatial-to-spectral harmonic-modulated array (SHA) is a MB-MIMO array that creates a mapping between different spatial directions and different `intermediate' frequency bins, enabling frequency-domain multiplexing. In an SHA, these intermediate frequencies, labeled $f_1, f_2, f_3$ in Fig.~\ref{fig:spatialspectralintro}(b) are harmonically related. When compared to a traditional beamforming  array, a single SHA can form multiple distinct beams and map the signal corresponding to each beam direction to a unique frequency. The fundamental benefit of SHAs when compared to traditional beamforming arrays is that they enable simultaneous beamforming toward multiple angles without the need for replicated or parallel beamforming hardware. In essence, the multiplexing that is traditionally achieved by hardware replication in traditional multi-beam beamforming arrays is instead accomplished by multiplexing in frequency via a spatial-to-spectral mapping. As a result, SHAs can outperform traditional multi-beam beamforming arrays in terms of power, cost and area.

As will be discussed throughout this article, SHAs have the potential to become key architectures for multiple 6G applications, including multi-user MIMO communications, joint communications and sensing (JCAS), and spatial filtering of in-band interferers.

Intuitively, SHAs can be understood by drawing a comparison to beam squint. In traditional phased arrays, beam squint occurs when the phase shifts intended to steer the beam at a particular frequency become misaligned at other frequencies, causing the beam to `squint' or shift direction. This is an unintended, frequency-dependent beam shift that becomes more pronounced with wider bandwidths \cite{beshary_spatial_2023}.

In contrast, spatial-to-spectral mapping intentionally uses frequency-dependent phase shifting to create multiple beams that point in different spatial directions for different frequencies. Instead of seeing beam direction shift as a limitation, SHAs deliberately utilize frequency-dependent phase-shifting to implement a spatial-to-spectral mapping. 
Note that in the arrays discussed in this article and the one shown in Fig.~\ref{fig:spatialspectralintro}(b) we assume that the effects of beam squint are negligible. However, depending on the signal bandwidth and the array size, this may not always be the case. Nevertheless, beam-squint mitigation techniques can be applied to SHAs similar to a traditional phased array.


This article provides an introduction to and overview of SHAs with a focus on MB-MIMO applications. Further, that understanding is extended by introducing two new SHA architectures and providing new analysis tools for such architectures. Namely, this article analyzes the effects of the harmonic-modulation waveform on SHA performance and introduces and analyzes the spatial-to-spectral degrees of freedom of several SHA architectures.

The article is organized as follows: Section \ref{section:intro} introduces the principle of operation of SHAs and an overview of existing and proposed architectures, Section \ref{section:lit} provides a literature review of existing SHA IC implementations and provides future directions for several MB-MIMO applications. Section \ref{section:example_tma} provides an analysis of the TMA to form an intuitive and mathematical foundation for SHAs. Section \ref{section:LOwaveform} builds on that foundation by analyzing the effects of the harmonic-modulation waveform on SHA performance. Section \ref{section:ssdofs} introduces the concept of spatial-to-spectral degrees of freedom and analyzes the degrees of freedom of two proposed SHA architectures. Lastly, conclusions are drawn in Section \ref{section:conclusion}. 

\section{Introduction to SHAs}
\label{section:intro}

 \begin{figure}[h]
    \centering
    \includegraphics[width=0.75\linewidth]{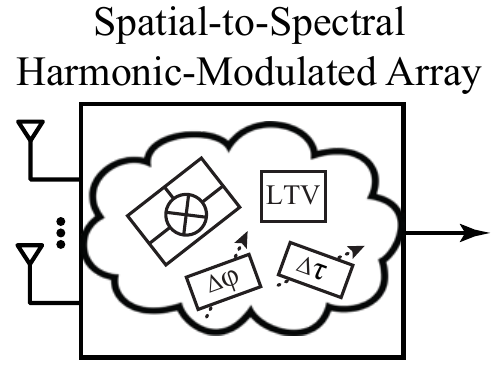}
    \caption{SHAs generally consist of LTV elements, such as mixers. Additionally, SHAs have a phase-frequency profile that can be determined by true-time delay elements and/or phase shifters.}
    \label{fig:ingredients}
\end{figure}

An SHA generally contains the following two key ingredients:
\begin{itemize}
    \item An element that `spreads' the signal information into several non-overlapping  frequency bands, such as a linear time-varying (LTV) element (i.e. mixer).  
    \item An element that imparts a frequency-dependent phase shift (i.e. true-time delay). 
\end{itemize}

As shown in Fig.~\ref{fig:ingredients}, an RF/mm-Wave SHA may consist of mixers, phase shifters and true-time delay (TTD) elements.

\begin{figure*}
    \centering
    \includegraphics[width=1\linewidth]{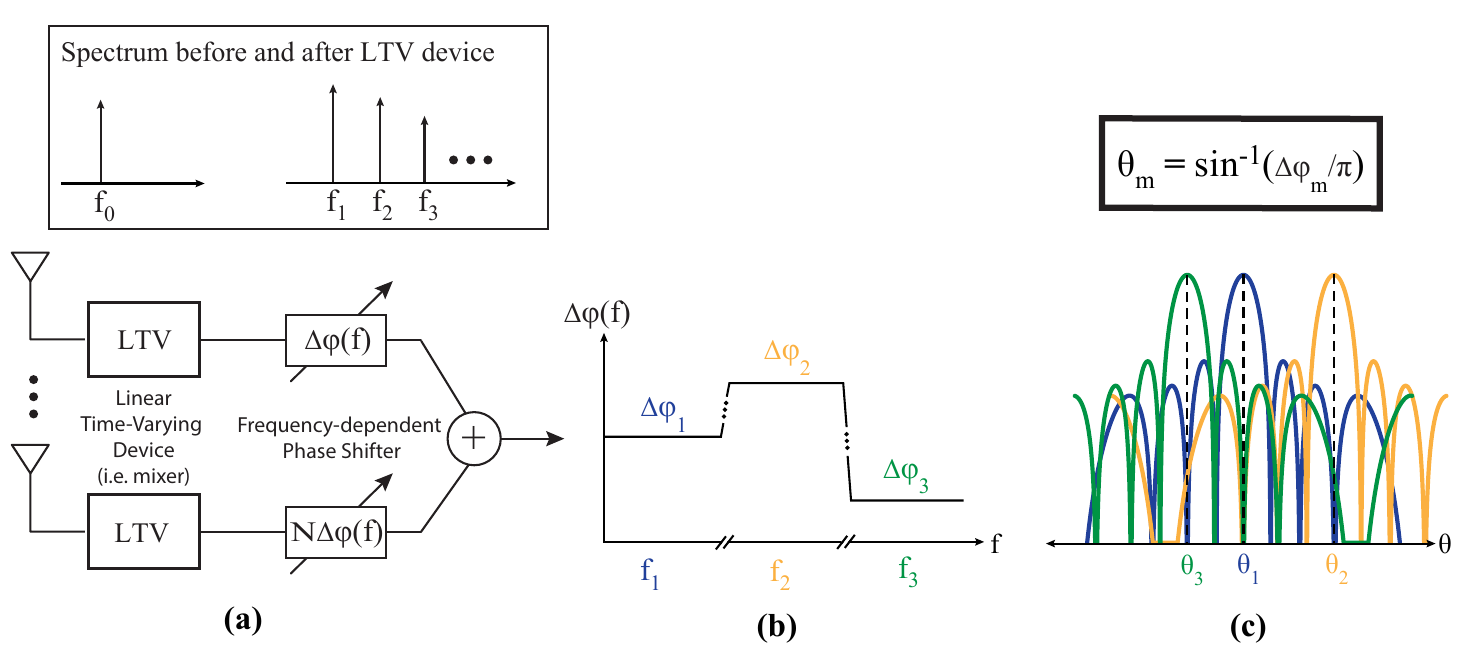}
    \caption{(a) A generic representation of an SHA with output frequencies $f_1, f_2, f_3$. (b) The SHA's phase-frequency relationship,  $\Delta\phi(f)$, is such that at $f_1, f_2, f_3$ the progressive phase shift across the array is $\Delta\phi_1, \Delta\phi_2, \Delta\phi_3$. (c) Due to the frequency-dependence of $\Delta\phi(f)$, a beam is formed at each frequency $f_1, f_2, f_3$ pointed toward angle $\theta_1, \theta_2, \theta_3$, respectively.}
    \label{fig:generic_ssa}
\end{figure*}

Fig.~\ref{fig:generic_ssa}(a) shows a generic SHA, consisting of an LTV element that `spreads' the input signal at $f_0$ to multiple harmonically related frequencies $f_1, f_2, f_3$. Additionally, the tunable frequency-dependent phase shifter provides an arbitrary phase shift $\Delta\phi_1, \Delta\phi_2, \Delta\phi_3$ at frequencies $f_1, f_2, f_3$, as shown in Fig.~\ref{fig:generic_ssa}(b). As a result of the distinct progressive phase shifts $\phi_m$ at each array element, the array beam patterns after signal combining are as shown in Fig.~\ref{fig:generic_ssa}(c). 
If we assume a half-wavelength antenna element spacing, the SHA creates a mapping between frequency $f_m$ and the beam centered at $\theta_m$. In other words, a receiving SHA would modulate a signal received at an angle-of-arrival $\theta_m$ to frequency $f_m$. Moreover, in a scenario where the effects of beam squint are negligible, such an SHA would be able to support wideband signals as well.

\begin{figure*}
    \centering
    \includegraphics[width=0.8\linewidth]{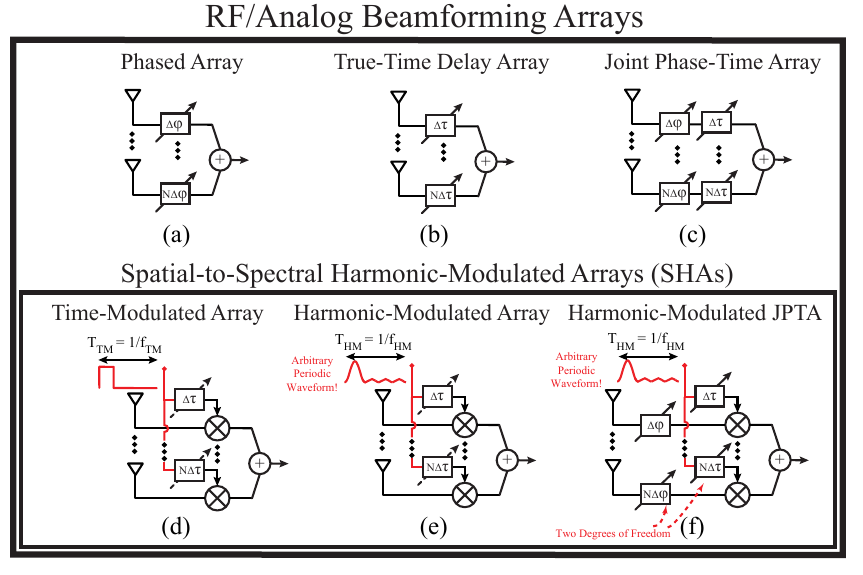}
    \caption{ (a) The phased array and (b) true-time delay array are two common beamforming architectures. Additionally, the (c) joint phase-time array has been previously explored to enable frequency-dependent beamforming. The remaining architectures are SHAs. (d) The time-modulated array is an example of an SHA where the waveform a square wave. In this paper, we generalize to (e) the harmonic-modulated array, where that waveform is an arbitrary periodic waveform. Additionally, this article explores architectures that provide multiple spatial-to-spectral degrees of freedom such as (f) the harmonic-modulated JPTA.}
    \label{fig:ssas_overview}
\end{figure*}

A categorized overview of analog and RF beamforming architectures is shown in Fig.~\ref{fig:ssas_overview}. These include the phased array, true-time delay array (TTDA) and joint-phase-time array (JPTA). One example of an SHA is the time-modulated array (TMA) shown in Fig.~\ref{fig:ssas_overview}(d). In a TMA, each channel is driven by non-overlapping square waves, such that at any point in time only one channel is conducting. This is equivalently modeled as a mixing stage with per-channel true-time delay elements. In this article, the harmonic-modulated array (HMA) is introduced. The HMA is a generalized version of the TMA where the mixer is instead driven by an arbitrary periodic waveform comprised of a fundamental tone at $f_{HM}$ and a set of harmonics. Additionally, this article introduces the concept of SHA spatial-to-spectral degrees of freedom and explores architectures like the the harmonic-modulated joint-phase-time array (HM-JPTA) which provides an additional degree of freedom. 

Importantly, an SHA differs functionally from a true-time delay array (TTDA) and a joint-phase-time array (JPTA), as highlighted in Fig.~\ref{fig:ssas_overview}. While the TTDA and JPTA create a mapping between frequency and space, they have relatively limited control of that mapping, As a result, they do not enable frequency-domain multiplexing for multi-beam applications to the same extent that SHAs do. 

Lastly, it is highlighted that the SHA mixer that creates the spatial-to-spectral mapping is distinct from the mixer to be used for up-conversion or down-conversion in a TX or RX channel, respectively. In this article, the mixer used to create the spatial-to-spectral mapping is referred to as the harmonic-modulation (HM) mixer and its corresponding LO as the harmonic-modulation LO (HM-LO). We emphasize that the term `harmonic-modulation' is used purposefully instead of `time-modulation' to highlight that SHAs do not strictly require a square-wave-driven mixer.

\section{SHA Applications and Literature Review}
\label{section:lit}
Certain SHAs have been the focus of recent literature across several disciplines. The TMA is an example of an SHA that has been extensively studied in the antenna propagation and signal processing communities \cite{rocca_4-d_2014,shanks_new_1961,poli_harmonic_2011,ni_high-efficiency_2023,yang_sideband_2002,kummer_ultra-low_1963}. While, RF/mm-Wave IC implementations of SHAs, including the TMA, remain relatively unexplored, a summary of SHA IC implementations is presented below. In particular, we focus on the key MB-MIMO applications that stand to benefit from the multi-beam capabilities of the SHA. These application spaces are 1) multi-user MIMO communications, 2) joint sensing and communications, 3) and spatial filtering of in-band interferers.

\subsection{SHA ICs for Multi-User MIMO Communications}
Future base stations are expected to contain MIMO antenna arrays with thousands of elements serving hundreds of users \cite{noauthor_nokia_nodate-1}. This demand for highly scalable multi-user beamforming arrays is well-aligned with the benefits of SHAs that have been outlined.

A four-channel 26-33 GHz RX time-modulated JPTA (TM-JPTA) is presented in \cite{huang_time-modulated_2023, ramkumar_time_2025}. The proposed architecture implements time-modulation with a tail-current switch at the LNA. Further, IF phase shifters at each channel enable the beams to be steered. The four-channel TM-JPTA forms five beams each mapped to a distinct non-overlapping IF frequency band for independent processing, enabling single-wire multi-user communications via spatial-to-spectral mapping. One drawback of the proposed architecture is that, by nature, time-modulation sacrifices array gain. 

Additionally, \cite{choi_d-band_2025} implements a four-channel D-band TX array forming up to 20 concurrent beams. This work introduces an architecture that cascades a traditional analog beamforming array with a TMA to effectively realize RF-domain beam number multiplication. Lastly, this work proposes a phase-modulation-based TMA to recover the array factor losses that a traditional TMA experiences.

SHA architectures employing multi-phase mixers have been demonstrated for applications that require many independently steerable beams \cite{ahasan_frequency-domain-multiplexing_2021, garg_28-ghz_2021}. Multi-phase mixers are capable of creating an arbitrary spatial-to-spectral mapping with a number of spatial-to-spectral degrees of freedom up to the number of channels $N$. However, this architecture suffers from several drawbacks. Firstly, the amplitude and phase resolution are limited by the amplitude resolution of the mixer weights and number of mixer phases. Moreover, the multi-phase mixer SHA requires significant hardware replication in a way that makes scaling such an architecture to large arrays a challenge.

\subsection{SHA ICs for Joint Communication and Sensing}

Joint communication and sensing (JCAS) is a highly desirable feature of future 6G systems as it enables systems with improved spectral and energy efficiency by allowing the reuse of the same radio resources (frequency, time, hardware) for two distinct purposes \cite{wild_joint_2021}. Further, fast localization is imperative for low-latency communications and sensing. It has been previously shown that arrays containing true-time delay elements enable so-called rainbow-beam training, offering substantially increased angle-of-arrival and angle-of-departure estimation speed \cite{wadaskar_3d_2021, lin_wideband_2022, xu_switching-less_2022, lin_multi-mode_2022, xu_ttd-based_2025, wadaskar_fast_2025}. The multi-beam capabilities of SHAs and the possibility for very fast localization of multiple receivers/transmitters make SHAs highly attractive for future 6G systems that support JCAS. However, to the authors knowledge, SHA ICs have yet to be demonstrated for JCAS applications.

Meanwhile, a JPTA-based 28 GHz transmitter array for JCAS is presented in \cite{mannem_reconfigurable_2023}. In this work, the combination of the tunable phase shift and tunable TTD between array elements enable the array to operate in three distinct modes: 1) as a MB-MIMO communication array, 2) as a communication array with variable security by means of tuning the phase and time delay elements, and 3) a localization (sensing) array. However, since JPTAs do not fully enable frequency-domain multiplexing (in the way that a HM-JPTA might), such an array does not operate as a communication and sensing array simultaneously.

A similar implementation that incorporates harmonic modulation (including time modulation) could enable an array that operates in two modes: 1) a localization mode, where the spatial-to-spectral mapping enables fast angle estimation and 2) a simultaneous communication and sensing (possibly radar) mode. Such an array would enable time, frequency and hardware resources to be shared for both communications and sensing. Importantly, the SHA would require at least two spatial-to-spectral degrees of freedom (two tunable phase elements) to enable the communication and sensing beams to be independently steered. The HM-JPTA architecture may be suitable.

\subsection{SHA ICs for Interference Cancellation via Spatial Filtering}
The dense deployment of small cells and the use of high-frequency bands in 6G increase the potential for various forms of interference, including co-channel interference. Additionally, the demand for flexible, low-latency, high-throughput wireless networks necessitates the use of largely digital beamforming MB-MIMO base stations with per-channel RF-to-digital signal chains. One key drawback of such architectures is that, in the presence of interference, the high dynamic range requirements of the per-channel ADCs lead to prohibitively high power and area costs. 

It has been demonstrated that the use of full-rank analog spatial filters significantly relaxes the ADC dynamic range requirement, reducing the ADC power consumption by up to 85\% \cite{van_den_heuvel_full_2012}. As a result, millimeter-wave IC spatial filtering front-ends for digital MB-MIMO have been implemented for both RX \cite{zhang_01--31ghz_2017,zhang_arbitrary_2017,zhang_23--29ghz_2022,huang_2337-ghz_2023,garg_28-ghz_2021,huang_full-fov_2019,eleraky_mm-wave_2024} and TX\cite{eleraky_mm-wave_2024,hu_28-ghz_2024}. However, most array-based spatial filtering implementations consist of traditional phase-shifter-based analog beamformers and, thus, suffer from poor scalability with increasing number of interference sources rejected. That is to say, a spatial filtering front-end that aims to place spatial nulls at $K$ angles is implemented with $K$ independent analog beamformers. 

In the same way that SHAs enable multi-user communication, SHAs can be utilized to implement scalable multi-null spatial filtering with minimal hardware replication. However, SHAs for spatial filtering are quite unexplored. A multi-phase mixer-based SHA that supports both beam and null steering is presented in \cite{garg_28-ghz_2021}. The multi-phase mixer approach enables a large number of spatial-to-spectral degrees of freedom but suffers from relatively coarse angle resolution. 

Importantly, we note that independent beam and null steerability is highly important for spatial filtering front-ends as the amount of interference rejection is highly dependent on accurate beam and null placement. Thus, spatial filtering front-ends employing SHAs will rely on multiple spatial-to-spectral degrees of freedom to ensure sufficient beam and null accuracy to simultaneously serve multiple users and reject interferers. Further, the resolution of phase-shifting elements is imperative in such applications, as null placement is generally very sensitive to amplitude and phase errors. 




\section{Guiding Example: The Time-Modulated Array}
\label{section:example_tma}
The time-modulated array, an example of an SHA, is reviewed in this section as a guiding example. By first gaining a mathematical and intuitive understanding of the TMA, we aim to generalize that understanding to other SHA architectures.

\begin{figure*}
    \centering
    \includegraphics[width=1\linewidth]{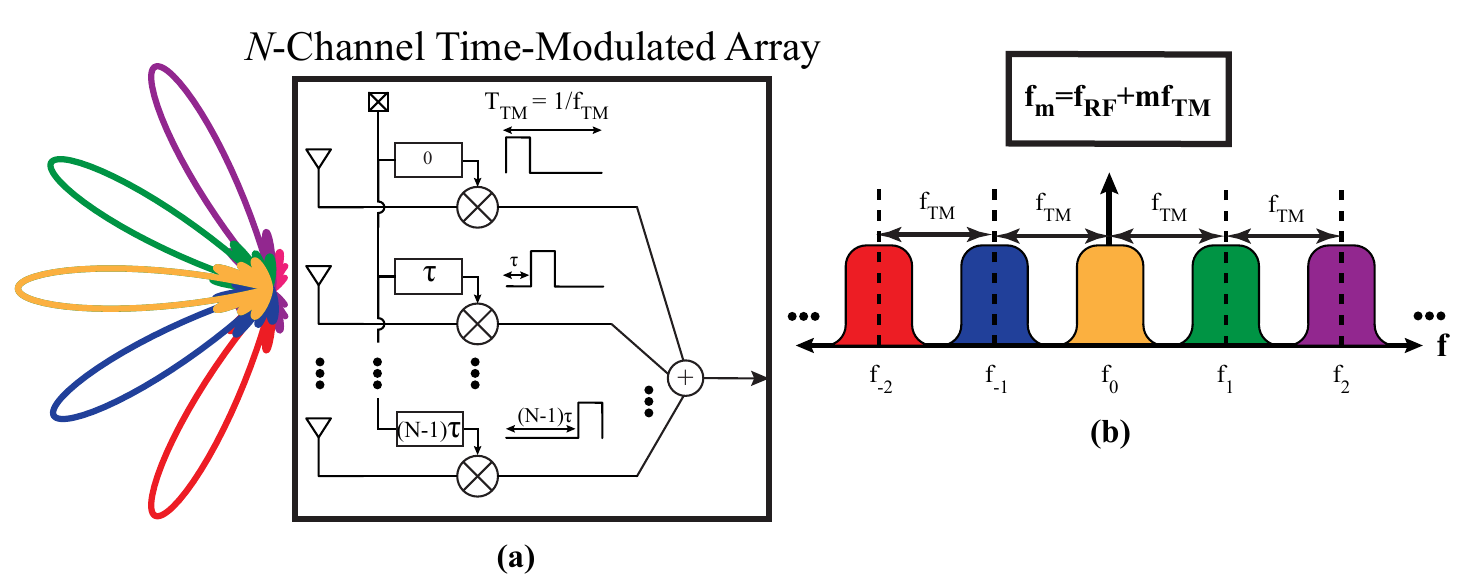}
    \caption{(a) A block diagram of an $N$-channel TMA. (b) The TMA output spectrum. At each frequency $f_m$, the TMA forms a beam centered at a unique angle $\theta_m$.}
    \label{fig:tma_intro}
\end{figure*}

Fig.~\ref{fig:tma_intro}(a) shows an $N$-channel TMA, characterized by an on-off switch, or mixer, at each channel. In the TMA, the switches are driven by non-overlapping single-ended square waves with fundamental period of $T_{TM}$, such that at any given time only one path is conducting. Equivalently, each mixer is driven by a square wave with a duty cycle of $1/N$ that is progressively time-delayed by $\tau=T_{TM}/N$ from channel to channel. The outputs of each mixer are combined to form the array output. 

The spectrum of the square wave HM-LO signal contains harmonic content at DC, the fundamental frequency $f_{TM}$ and at certain harmonics of $f_{TM}$, depending on the exact shape of the square wave. At the output of the mixer, the received signal at each array channel has been spread to many  frequencies $f_{IF,k} = f_{RF} \pm kf_{TM}$ for $k \in \{0, 1, 2, 3, ...\}$. Due to the double-sided nature of the mixer's output spectrum, we can also represent this set of frequencies as  $f_{IF,m} = f_{RF} + mf_{TM}$ for $m \in \{..,-2, -1, 0, 1, 2, ..\}$.  Additionally, at each frequency $f_{IF,m}$ the combination of the true-time delay and mixer impart a frequency-dependent phase-shift $\phi_{m} = 2\pi(mf_{TM})\tau$. It is noted that $\phi_{m}$ is negative for lower side-bands (i.e. for negative $m$). 

By making this observation, the TMA's principle of operation can be summarized. The true-time delay imparts a frequency-dependent phase shift to each of the square wave's $k$ harmonics. Then, the mixer takes the sum and difference of the RF frequency and each of those harmonics, as well as the sum and difference of phases. It follows that at each of the output frequencies $f_{IF,m}$ the signal is imparted a unique phase $\phi_{m}$. At each channel $n$, the TMA has a progressive time-delay of $n\tau$. For a given value of $\tau$ the array will create a mapping between different angles of arrival, $\theta_m$, and the different output frequencies $f_{IF,m}$. While not shown in the TMA of Fig.~\ref{fig:tma_intro}, adding tunability to the TTD elements enables the angular distance between adjacent beams to be controlled. A rigorous analysis of the TMA principle of operation is presented in \cite{huang_time-modulated_2023}.

\section{Harmonic-Modulation LO Waveform Shaping}
\label{section:LOwaveform}

In this article, we draw the distinction between `time-modulated' and `harmonic-modulated' arrays because the shape of the the harmonic modulation LO waveform in SHAs highly impacts performance metrics including gain, noise and bandwidth. As a result, the practicality and feasibility of the SHA implementation is highly determined by this waveform.

The TMA is one possible implementation of an SHA where the modulation waveform is a square wave. However, a generic SHA can have an arbitrary periodic HM-LO waveform, $a(t)$. We can write $a(t)$ in terms of the complex Fourier series

\begin{equation}
a(t) = \sum_{m=-\infty} ^{\infty}\beta_{m} e^{j2\pi mf_{HM}t}.
\end{equation}

 The HM-LO waveform, $a(t)$, is characterized by the following parameters: 1) the HM-LO fundamental frequency, $f_{HM}$ and 2) the HM-LO Fourier series coefficients, $\beta_m$, at each harmonic $m$. Further, let's define a term $M$, the highest value of $m$ for which the Fourier series has a non-zero coefficient. We can then rewrite the complex Fourier series of $a(t)$ as 
 
\begin{equation}
a(t) = \sum_{m=-M} ^{M}\beta_{m} e^{j2\pi mf_{HM}t}.
\end{equation}

\subsection{HM-LO Waveform Effects on Bandwidth}
The fundamental frequency, $f_{HM}$, of the HM-LO waveform determines the spacing between each spatial-to-spectral frequency band as is illustrated in Fig.~\ref{fig:tma_intro}(b) for the TMA. Specifically, $f_{HM} > f_{BW}$, where $f_{BW}$ is the RF channel bandwidth, to ensure non-overlapping frequency bands \cite{huang_time-modulated_2023}. Further, as discussed in \cite{ramkumar_time_2025}, to provide an additional transition bandwidth, $f_{TR}$, the requirement on $f_{HM}$ is $f_{HM} > f_{BW} + f_{TR}$. 

Additionally, the maximum HM-LO harmonic of interest, $M$,  determines the bandwidth that the circuits at the HM-LO and HM mixer output ports must support. For example, for an HMA driven by an HM-LO waveform with spectral energy at $DC, f_{HM}, 2f_{HM}, ..., Mf_{HM}$, the HM-LO distribution network must have a bandwidth greater than $Mf_{HM}$. Similarly, any circuits at the output of the HM mixer, including the combining or splitting network, must have a bandwidth supporting all harmonic bands ($m = -M$ to $m = M$). As a result, designing the HM-LO waveform in a way that minimizes $M$ can relax bandwidth requirements for the SHA circuits.


\subsection{HM-LO Waveform Effects on Array Signal Power, Noise Power and Signal-to-Noise Ratio}
As described in \cite{huang_time-modulated_2023} a TMA provides, at most, a gain of $0~dB$. Moreover, the noise power at the output of an $N$-channel TMA is lower than at the individual channel such that the array maintains a signal-to-noise ratio (SNR) boost, or array gain, of $10\times log_{10}(N)$ in $dB$. For a generic SHA with an arbitrary HM-LO it can be shown that there exists a tradeoff between gain and the number of HM-LO harmonics.

Consider the four-channel SHA shown in Fig.~\ref{fig:hmspectra}(a). Here, it is assumed that the frequency-dependent phase shifting occurs at the HM-LO port. The $n$-th mixer's HM-LO port is driven by $a_n(t)$, a phase and/or time-delayed copy of $a(t)$, which can be represented by the following complex Fourier series at each channel $n$.  

\begin{equation}
a_n(t) = \sum_{m=-\infty} ^{\infty}\beta_{n,m} e^{j2\pi mf_{HM}t}
\end{equation}

Neglecting the loss of the phase shifting elements for simplicity, only the phase of $a_n(t)$ varies with $n$. Equivalently, we can write that the magnitude of the Fourier coefficients is the same for all channels $n$:

\begin{equation}
|\beta_{n,m}| = |\beta_m|.
\label{eqn:betas}
\end{equation}

\subsubsection{Signal Power Gain}
By nature, the SHA has a harmonic-dependent signal power gain that is dependent on $\beta_m$. Additionally, the signal of interest is correlated across channels and sums coherently in voltage. Thus, the signal power gain of such an array if we assume lossless combining is 
\begin{equation}
G_{sig,m} = \left|\sum_{n=0} ^{N-1}|\beta_{n,m}|\right|^2.
\end{equation}

If we consider the simplification shown in Eqn.~\ref{eqn:betas}, we can show that at all harmonics $m$ the corresponding spatial beam provides a signal power gain of
\begin{equation}
G_{sig,m} = \left|\sum_{n=0} ^{N-1}|\beta_{m}|\right|^2 = N^2|\beta_m|^2.
\label{eqn:gsig}
\end{equation}

\subsubsection{Noise Power Gain}
It can similarly be shown that the SHA provides a harmonic-dependent noise power gain. In an SHA, the uncorrelated noise at each channel sums coherently in power. As a result, we can write that the SHA provides a harmonic-dependent noise power gain of 

\begin{equation}
G_{noise,m} =\sum_{n=0} ^{N-1}|\beta_{m}|^2 = N|\beta_{m}|^2.
\label{eqn:gnoise}
\end{equation}

\subsubsection{Signal-to-Noise Ratio and Array Gain}
The array's improvement of signal-to-noise ratio is referred to as the array gain. By combining Eqn.~\ref{eqn:gsig} and Eqn.~\ref{eqn:gnoise} we can compute the array gain at each harmonic $m$.

\begin{equation}
AG_m = \frac{SNR_{out,m}}{SNR_{in}} = \frac{G_{sig,m}}{G_{noise,m}} = N.
\label{eqn:AGm}
\end{equation}

Eqn.~\ref{eqn:AGm} shows that that for any arbitrary HM waveform the SHA provides an array gain of $N$, or $10\times log_{10}(N)$ in $dB$, at all harmonics $m$.

\subsubsection{Harmonic Loss in SHAs with Many HM-LO Harmonics}
\begin{figure}
    \centering
    \includegraphics[width=1\linewidth]{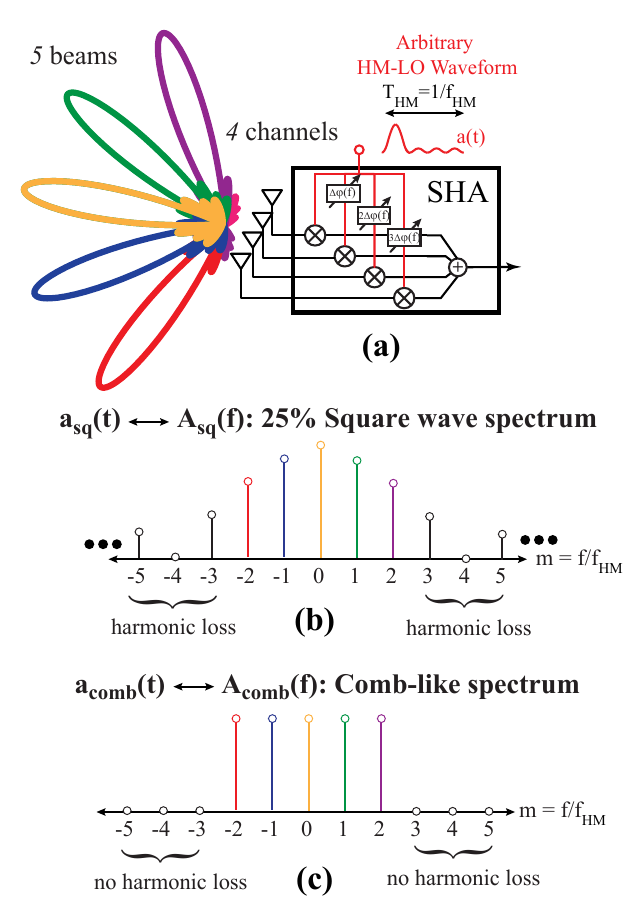}
    \caption{(a) A block diagram of a four-channel, five-beam SHA driven by an arbitrary HM-LO waveform. (b) The magnitude of the square wave spectrum. The square wave is one possible candidate but it exhibits harmonic loss as some of the spectral energy of the signal is not mapped to any beam. (c) The magnitude of the comb-like spectrum. A waveform whose spectrum is comb-like redistributes the unused spectral energy to the harmonics that corresponds to the SHA's beams, such that none of the HM-LO's spectral energy is gone to waste.}
    \label{fig:hmspectra}
\end{figure}

This section aims to shed light on the tradeoff between the number of HM-LO harmonics and SHA signal power gain. While the number of beams can vary depending on application, we assume as in \cite{huang_time-modulated_2023} that the number of beams $K$ is equal to $N+1$. Let's consider the four-channel, five-beam SHA shown in Fig.~\ref{fig:hmspectra}(a). Two possible HM-LO waveforms for this SHA whose spectra are shown in Fig.~\ref{fig:hmspectra}(b)(c) are discussed. The HM-LO spectrum shown in Fig.~\ref{fig:hmspectra}(b) corresponds to a square wave, $a_{sq}(t)$ switching between 0 and 1 with a duty cycle of $1/4$. This is the typical HM-LO waveform in a four-channel TMA. The TMA creates a spatial-to-spectral mapping between those five beams and the five output frequencies $f_m = f_{RF} +mf_{HM}$ for $m\in[-2, -1, 0, 1, 2]$. For all other harmonics of the HM-LO square wave, the spectral power of the HM-LO signal is unused. We refer to this as harmonic loss. One can imagine that redistributing the unused spectral power to the harmonics of the HM-LO waveform that correspond to the five beams would increase the signal power gain $G_{sig,m}$ at the harmonics of interest. Indeed, this highlights that there exists a tradeoff between the number of HM-LO harmonics and the achievable signal power gain for a fixed HM-LO power. 

Fig.~\ref{fig:hmspectra}(c) shows the spectrum of an HM-LO waveform, $a_{comb}(t)$, that we refer to as comb-like since its power spectral density consists of equally spaced equal-height peaks at each harmonic. An SHA driven by $a_{comb}(t)$ would exhibit the following potentially desirable properties 1) no harmonic loss and 2) uniform signal power gain for all beams ($G_{sig,0}=G_{sig,1}= ... = G_{sig}$). 

\begin{figure}
    \centering
    \includegraphics[width=1\linewidth]{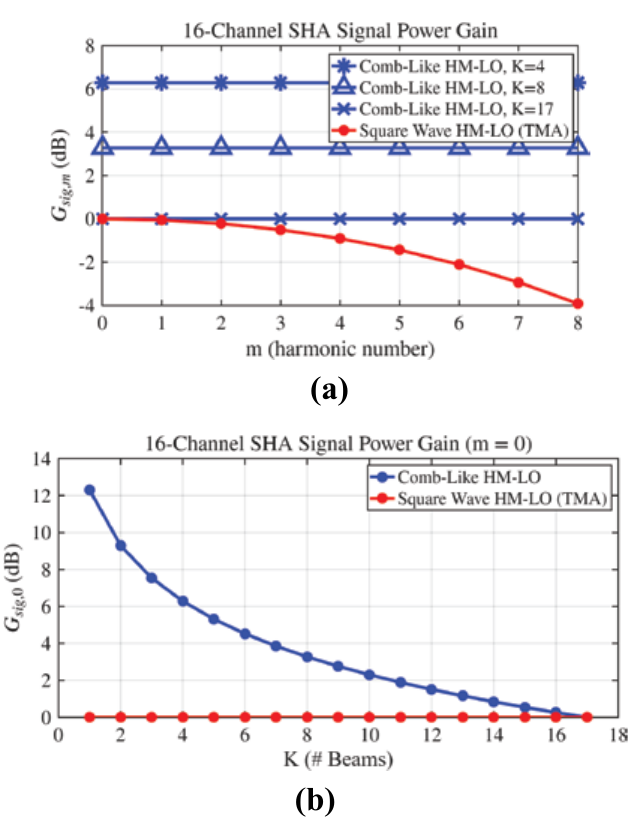}
    \caption{(a) SHA gain vs. harmonic number $m$ for square and comb-like HM-LO waveforms. When driven by a square-wave, the SHA gain depends on the harmonic $m$. Meanwhile, an SHA driven by a comb-like waveform provides equal gain for all beams. (b) SHA gain ($m = 0$) vs. K ($\#$ Beams). In systems with few beams the comb-like HM-LO waveform enables the SHA to have higher signal power gain than the TMA. This is because the TMA exhibits more harmonic loss when there are fewer beams.}
    \label{fig:gainvsk}
\end{figure}

Fig.~\ref{fig:gainvsk} shows the signal power gain at each harmonic, $G_{sig,m}$, for a sixteen-channel SHA. Specifically, the square wave and comb-like HM-LO waveform are compared while maintaining fixed HM-LO power $\left( \sum_{m=-\infty} ^{\infty}|\beta_{m}|^2\right)$. Fig.~\ref{fig:gainvsk}(a) shows that the square-wave-driven SHA (TMA) has decreasing gain for increasing $m$. Additionally, it can be observed that the comb-like SHA has uniform gain for all harmonics. 

As shown in Fig.~\ref{fig:gainvsk}(b), using a comb-like waveform instead of a square wave can improve the signal power gain and this effect is more pronounced for smaller $K$. This is because the TMA exhibits more harmonic loss when there are fewer beams as more of the square wave's spectral energy is unused. 

In summary, it has been shown that controlling the number of HM-LO harmonics and the magnitude of the complex Fourier series coefficients at those harmonics can minimize harmonic loss and improve signal power gain while maintaining a fixed HM-LO power. 

\section{Spatial-to-Spectral Degrees of Freedom}
\label{section:ssdofs}

\begin{figure*}
    \centering
    \includegraphics[width=1\linewidth]{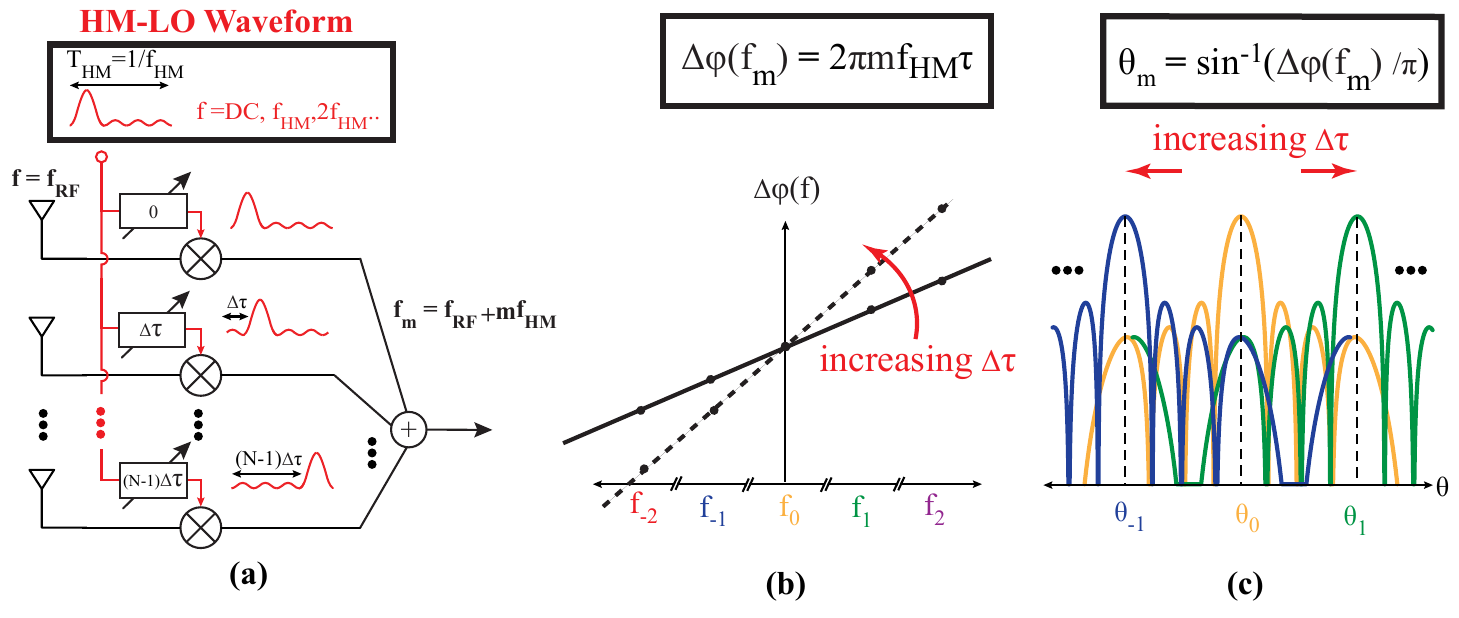}
    \caption{(a) A block diagram of an $N$-channel HMA with one spatial-to-spectral degree of freedom. (b) The array's phase-frequency profile, $\Delta\phi(f)$, which varies as $\Delta\tau$ is tuned. (c) The HMA spatial-to-spectral beams. The beam at $\theta_m$ maps to frequency $f_m$. When $\Delta\tau$ varies, the angular spacing between the beams varies, providing a spatial-to-spectral degree of freedom.}
    \label{fig:hma_one_dof}
\end{figure*}

Fundamentally, an SHA is constrained in its ability to create an arbitrary spatial-to-spectral mapping by the constraints on the array channel's phase-frequency relationship. For example, in the TMA shown in Fig.~\ref{fig:tma_intro}(a), where the phase-frequency relationship is strictly linear due to the nature of the true-time delay, the fixed linear relationship means that the formed beams are static. While the progressive time-delay $\tau$ in a TMA can be made tunable, such as in the HMA of Fig.~\ref{fig:hma_one_dof}(a), this still only enables a single degree of freedom in the spatial-to-spectral mapping. 


For an arbitrary SHA, let's call $\Delta\phi(f)$ the array's progressive (channel-to-channel) phase shift as a function of SHA output frequency, $f$. In an ideal case, such as the generic SHA shown in Fig.~\ref{fig:generic_ssa}, $\Delta\phi(f)$ can be arbitrarily programmed at each frequency of interest. Fig.~\ref{fig:hma_one_dof} shows $\Delta\phi(f_m) = 2\pi mf_{HM}\Delta\tau$ for an HMA as the value of $\tau$ is varied. Importantly, with a single degree of freedom, $\Delta\tau$, independent beam steering can only be achieved for a single beam.
We note that an alternative implementation of an SHA with a single spatial-to-spectral degree of freedom is presented in \cite{huang_time-modulated_2023}, where the time-delay element is static and the degree of freedom is provided by a downstream tunable phase shifter.

\subsection{SHAs with Two Spatial-to-Spectral Degrees of Freedom}

\begin{figure*}
    \centering
    \includegraphics[width=1\linewidth]{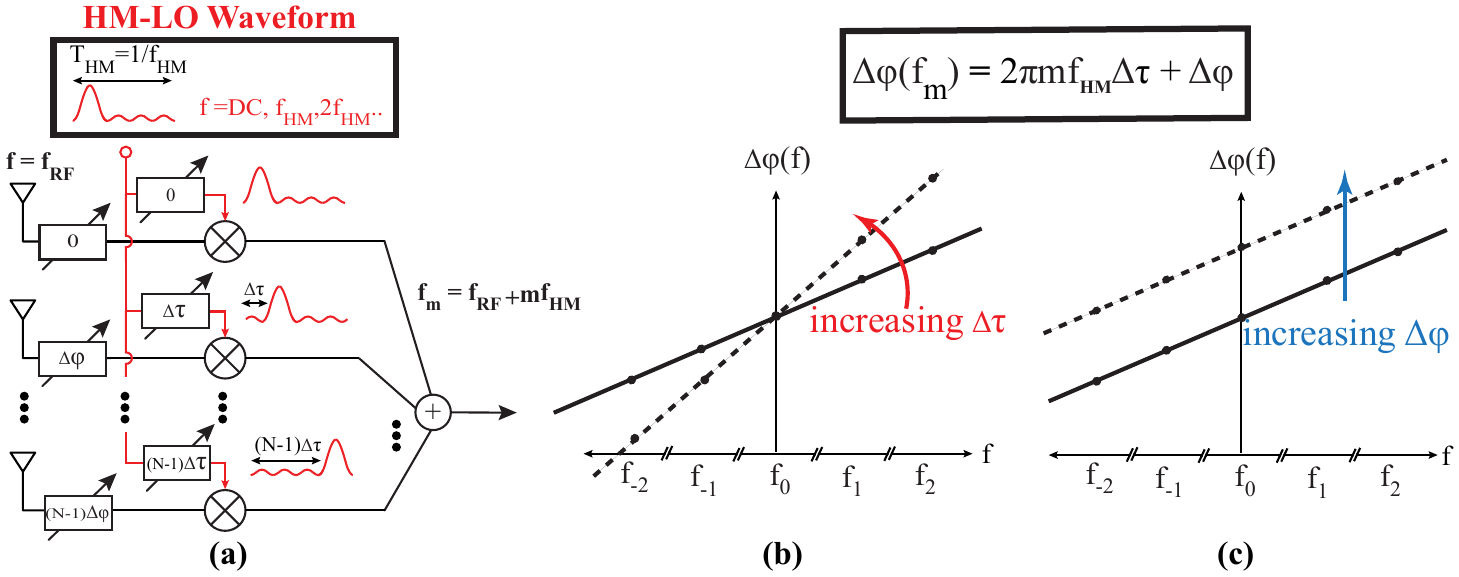}
    \caption{(a) The block diagram of an $N$-channel HM-JPTA which has two spatial-to-spectral degrees of freedom. (b) The phase-frequency relationship, $\Delta\phi(f)$, changes with $\Delta\tau$ to providing the first spatial-to-spectral degree of freedom. (c) The tunable phase shifter provides a second independent spatial-to-spectral degree of freedom by adding the same phase at all frequencies $f_m$.}
    \label{fig:hmjpta_two_dof}
\end{figure*}

One straightforward way to add a second spatial-to-spectral degree of freedom is by taking the HMA architecture shown in Fig.~\ref{fig:hma_one_dof}(a) and adding an additional phase shifter, often in the RF path, as shown in Fig.~\ref{fig:hmjpta_two_dof}(a). We refer to such an SHA as a harmonic-modulated joint phase-time array (HM-JPTA). Note that the term harmonic-modulated, as opposed to time-modulated, highlights that the HM-JPTA does not strictly require a square-wave-driven mixer. The HM-JPTA's phase-frequency relationship can be written as:

\begin{equation}
\Delta\phi(f_m) =  2\pi mf_{HM}\Delta\tau + \Delta\phi
\label{eqn:two_dof}
\end{equation}

Eqn.~\ref{eqn:two_dof} can be reformulated as a matrix expression of the form $\boldsymbol{y}=\boldsymbol{A}\boldsymbol{x}$ as follows:
\begin{gather}
 \begin{bmatrix} 
 \vdots \\
 \Delta\phi(f_{-2}) \\
 \Delta\phi(f_{-1}) \\
 \Delta\phi(f_0) \\
 \Delta\phi(f_1) \\
 \Delta\phi(f_2) \\
 \vdots
 \end{bmatrix}
 =
 \begin{bmatrix} 
 \vdots &  \vdots \\
1 & -2 \\
1 & -1 \\
1 & 0 \\
1 & 1 \\
1 & 2 \\
 \vdots  & \vdots 
 \end{bmatrix}
\begin{bmatrix} 
 \Delta\phi \\
 2\pi f_{HM}\Delta\tau 
\end{bmatrix}
\label{eqn:mtx_two_dof}
\end{gather}

It can be seen that, for certain pairs of rows, a rank-two  matrix can be formed. For example: 

\begin{gather}
 \begin{bmatrix} 
 \Delta\phi(f_0) \\
 \Delta\phi(f_1) \\
 \end{bmatrix}
 =
 \begin{bmatrix} 
1 & 0 \\
1 & 1 \\
 \end{bmatrix}
\begin{bmatrix} 
 \Delta\phi \\
 2\pi f_{HM}\Delta\tau 
\end{bmatrix}
\label{eqn:mtx_two_dof_simp} 
\end{gather}

This rank-two matrix indicates that $\Delta\phi(f_0)$ and $\Delta\phi_(f_1)$ can be independently set by controlling the two independent variables $\Delta\phi$ and $ 2\pi f_{HM}\Delta\tau$ which directly map to tunable hardware elements. Thus, it can be concluded that the HM-JPTA in Fig.~\ref{fig:hmjpta_two_dof} is able to independently steer two spatial-to-spectral beams, say at frequencies $f_0, f_1$. Beams at additional harmonics ($f_2, f_3$...) will be formed depending on the harmonic nature of the HM-LO waveform. However, those beams will not be steered independently of the beams at $f_0, f_1$.

\begin{figure*}
    \centering
    \includegraphics[width=1\linewidth]{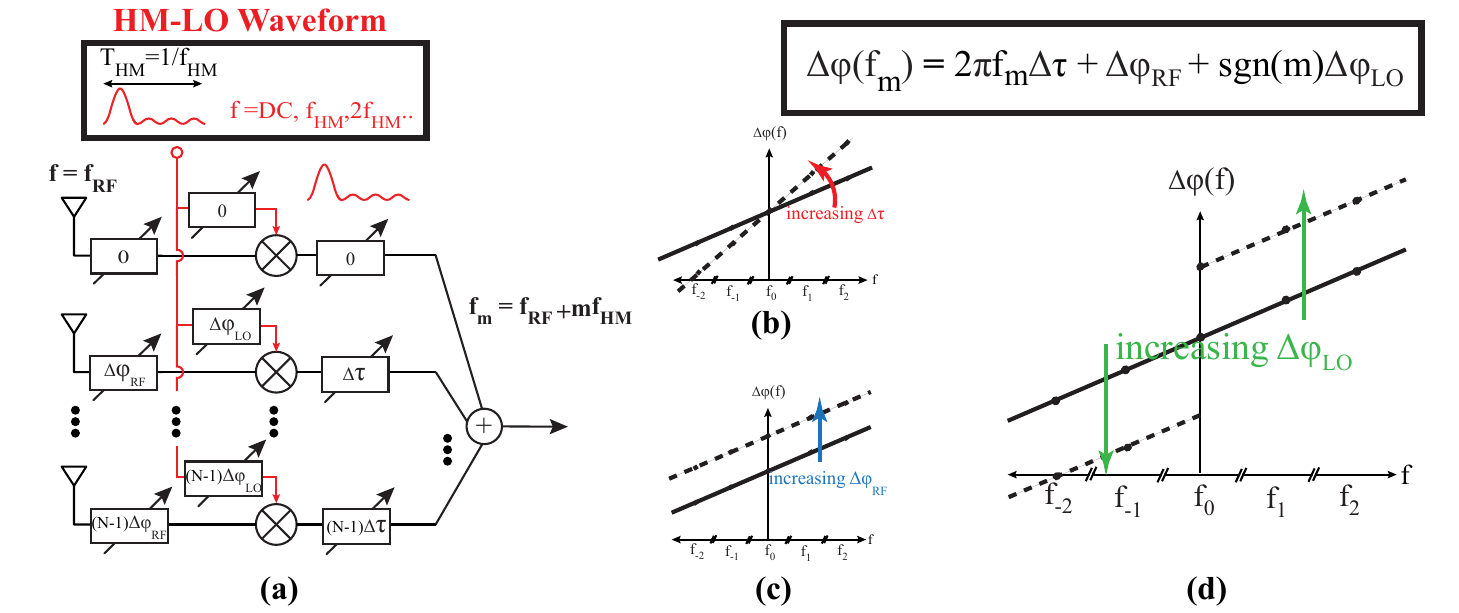}
    \caption{(a) The block diagram of an $N$-channel HM-JPTA with three spatial-to-spectral degrees of freedom. This architecture contains a wideband phase-shifter at the RF and HM-LO ports and a TTD element at the HM mixer output. (b) The tunable TTD element $\Delta\tau$ and (c) tunable RF phase shifter, $\Delta\phi_{RF}$ each provide one spatial-to-spectral degree of freedom. (d) The tunable phase shifter at the HM-LO, $\Delta\phi_{LO}$, provides the third spatial-to-spectral degree of freedom by imparting zero phase at $f_0$, positive phase for $f_m$ with positive $m$ and negative phase for $f_m$ with negative m.}
    \label{fig:hmjpta_three_dof}
\end{figure*}

\subsection{SHAs with Three Spatial-to-Spectral Degrees of Freedom}
Now consider the HM-JPTA shown in Fig.~\ref{fig:hmjpta_three_dof}(a). Compared to the HM-JPTA described earlier and shown in Fig.~\ref{fig:hmjpta_two_dof}(a), this architecture places the TTD element at the output of the HM mixer rather than the HM-LO port. Additionally, a wideband phase shifter is added at the mixer's HM-LO port.

Note that, since the TTD element is placed at the HM mixer's output path, it now imparts the following total phase:

\begin{equation}
\Delta\phi_\tau(f_m) = 2\pi(f_{RF}+mf_{HM})\Delta\tau
\label{eqn:three_dof_phi_f}
\end{equation}

The phase imparted by this TTD element contains two terms. The first term, $2\pi f_{RF}\Delta\tau$, is independent of $m$ and depends only on $f_{RF}$. The second term is identical to the phase imparted by a TTD at the HM-LO port, as in Eqn.~\ref{eqn:two_dof} and depends on $m$.

It follows that this architecture's corresponding phase-frequency relationship can then be written as:

\begin{equation}\
\Delta\phi(f_m) =  2\pi(f_{RF}+mf_{HM})\Delta\tau + \Delta\phi_{RF} + sgn(m)\Delta\phi_{LO}
\label{eqn:three_dof}
\end{equation}
where 
\begin{equation}\
sgn(m) =
\begin{cases} 
0 & m = 0 \\
1 & m > 0 \\
-1 & m < 0 \\
\end{cases}
\end{equation}

Note that for the harmonic $m$ = 0 the phase shifter corresponding to $\Delta\phi_{LO}$ does not impart a phase shift since the HM-LO signal is at DC. 

Eqn.~\ref{eqn:three_dof} can be reformulated as a matrix expression of the form $\boldsymbol{y}=\boldsymbol{A}\boldsymbol{x}$ as follows:
\begin{gather}
 \begin{bmatrix} 
 \vdots \\
 \Delta\phi(f_{-2}) \\
 \Delta\phi(f_{-1}) \\
 \Delta\phi(f_0) \\
 \Delta\phi(f_1) \\
 \Delta\phi(f_2) \\
 \vdots
 \end{bmatrix}
 =
 \begin{bmatrix} 
 \vdots & \cdots & \vdots \\
1 & -1 & -2\\
1 & -1 & -1\\
1 & 0 & 0\\
1 & 1 & 1\\
1 & 1 & 2\\
 \vdots & \cdots & \vdots 
 \end{bmatrix}
\begin{bmatrix} 
 \Delta\phi_{RF} + 2\pi f_{RF} \Delta\tau  \\
 \Delta\phi_{LO} \\
2\pi f_{HM} \Delta\tau
\end{bmatrix}
\label{eqn:mtx_three_dof}
\end{gather}

where 
\begin{gather}
 \boldsymbol{x}
 =
\begin{bmatrix} 
 \Delta\phi_{RF} + 2\pi f_{RF} \Delta\tau  \\
 \Delta\phi_{LO} \\
2\pi f_{HM} \Delta\tau
\end{bmatrix}
\end{gather}

Note that in Eqn.~\ref{eqn:mtx_three_dof} we combine the terms that are independent of $m$, namely $\Delta\phi_{RF}$ and the first term of Eqn.~$\ref{eqn:three_dof_phi_f}$, $2\pi f_{RF}\Delta\tau$. 

It can be seen that, for certain groups of rows, a rank-three  matrix can be formed. For example: 

\begin{gather}
 \begin{bmatrix} 
 \Delta\phi(f_0) \\
 \Delta\phi(f_1) \\
  \Delta\phi(f_2) \\
 \end{bmatrix}
 =
 \begin{bmatrix} 
1 & 0 & 0\\
1 & 1 & 1\\
1 & 1 & 2\\
 \end{bmatrix}
\begin{bmatrix} 
 \Delta\phi_{RF} + 2\pi f_{RF} \Delta\tau  \\
 \Delta\phi_{LO} \\
2\pi f_{HM} \Delta\tau
\end{bmatrix}
\label{eqn:mtx_three_dof_simp}
\end{gather}


While Eqn.~\ref{eqn:mtx_three_dof_simp} does form a rank-three matrix, it is not immediately obvious that the rank of three indicates that this architecture supports three independently steerable spatial-to-spectral beams. This is because the terms representing the independent variables (the terms in vector $\boldsymbol{x}$) do not uniquely correspond to tunable hardware elements as they did in Eqn.~\ref{eqn:mtx_two_dof_simp}.

However, we can consider the following elementary transformation matrix:
\begin{gather}
\boldsymbol{E}
=
\begin{bmatrix} 
1 & 0 & -f_{RF}/f_{HM}\\
0 & 1 & 0\\
0 & 0 & 1\\
\end{bmatrix}
\label{eqn:elementary_transformation_mtx}
\end{gather}

Further, we can define a new set of independent variables:

\begin{gather}
\boldsymbol{x'} = \boldsymbol{Ex} = 
 \begin{bmatrix} 
 \Delta\phi_{RF} \\
 \Delta\phi_{LO} \\
2\pi f_{HM} \Delta\tau
 \end{bmatrix}
\end{gather}

Thus, for any full-rank matrix $\boldsymbol{A}$, an equivalent matrix expression of the form $\boldsymbol{y}=\boldsymbol{A}\boldsymbol{E^{-1}}\boldsymbol{x'} = \boldsymbol{B}\boldsymbol{x'}$ can be written. Since $\boldsymbol{E}$ is an elementary matrix it is full-rank and invertible. As a result, $\boldsymbol{E^{-1}}$ is as well. Thus, it preserves the rank of $\boldsymbol{A}$ upon multiplication. As a result, $\boldsymbol{B}$ is full-rank itself.

It follows that Eqn.~\ref{eqn:mtx_three_dof_simp} can be rewritten in the form $\boldsymbol{y}=\boldsymbol{A}\boldsymbol{E^{-1}}\boldsymbol{x'} = \boldsymbol{B}\boldsymbol{x'}$ revealing that a matrix $\boldsymbol{A}$ of rank three is indicative of three distinct spatial-to-spectral degrees of freedom that can each be controlled by the independently tunable phase and TTD elements. As an example, for the SHA shown in Fig.~\ref{fig:hmjpta_three_dof}(a), the beams formed at frequencies $f_0, f_1, f_2$ are independently steerable, or orthogonal. 

\subsection{SHAs with $>$ 3 Spatial-to-Spectral Degrees of Freedom}


One of the primary limitations of the SHAs discussed so far is that they can, at most, independently steer two or three spatial-to-spectral beams. For some applications, it may be desirable to have more than three independently steerable beams. While relatively unexplored, two SHA architectures that can, in theory, support an arbitrarily large number of spatial-to-spectral degrees of freedom are discussed below. However, tunability comes at a cost as these architecture require substantial per-channel hardware replication.

\subsubsection{Multi-phase Mixers}
A multi-phase mixer-based SHA was implemented in \cite{garg_28-ghz_2021}. Each receiver channel contains a mixer followed by an IF multi-phase mixing stage. Each channel's IF multi-phase mixer consists of $M$ sub-paths each containing a variable-gain amplifier and time-delayed square-wave driven mixer. The VGA amplitude weights at each sub-path can be used to generate an arbitrary amplitude and phase at each array channel for each IF frequency. One way to intuitively understand the multi-phase mixer is by comparing it to a harmonic-reject mixer, where parallel amplitude-weighted mixers can form an equivalent mixer that nulls the third and fifth harmonics, for example \cite{weldon_175-ghz_2001}. The multi-phase mixer combines that same core idea with a time-delayed square wave LO to create a spatial-to-spectral mapping. Multi-phase mixers are capable of creating an arbitrary spatial-to-spectral mapping with a number of spatial-to-spectral degrees of freedom up to the number of channels. However, this architecture suffers from several key drawbacks. Firstly, the amplitude and phase resolution is limited by the amplitude resolution of the VGAs and number of mixer phases. Additionally, the multi-phase-mixer-based SHA requires significant hardware replication making scaling a challenge.

\subsubsection{Narrowband Phase Shifters}
One seemingly unexplored SHA architecture is discussed below. It utilizes multiple parallel narrowband phase shifters at the mixer's HM-LO port to independently set the progressive array phase shift uniquely at each frequency of interest. Such an architecture would enable an SHA with a number of spatial-to-spectral degrees of freedom equal to the number of parallel phase shifters in each channel. Theoretically, if the phase shifters have a sufficiently steep rolloff, spatial-to-spectral beamforming with sufficient isolation should be possible. However, the roll-off of the phase shifter response will limit performance.

\section{Conclusion}
\label{section:conclusion}
This article provided an overview of different spatial-to-spectral harmonic-modulated array architectures and presented a summary of existing IC implementations for various MB-MIMO applications including multi-user communications, joint communication and sensing, and spatial filtering for interference cancellation. Additionally, this article introduced the concept of a harmonic-modulated array and provided a detailed analysis of the effects of the HM-LO waveform on key performance metrics including gain, noise and bandwidth. Lastly, an analysis of SHA spatial-to-spectral degrees of freedom is provided and architectures with two and three spatial-to-spectral degrees of freedom are introduced. In doing so, this article provides key insights and guidelines for designing SHAs for 6G MB-MIMO applications.

\section{Acknowledgments}
This work was supported in part by the Center for Ubiquitous Connectivity (CUbiC), sponsored by Semiconductor Research Corporation (SRC) and Defense Advanced Research Projects Agency (DARPA) under the JUMP 2.0 program. This material is based upon work supported
by the National Science Foundation Graduate Research Fellowship under Grant No 2146752. Any opinion, findings, and conclusions or recommendations expressed in this material are
those of the authors(s) and do not necessarily reflect the views of the National Science
Foundation.

\bibliographystyle{ieeetr} 
\bibliography{bibtex/bib/Spatial_Spectral_Overview_PaperVersion2}

\end{document}